\documentclass{article}
\usepackage[utf8]{inputenc}
\usepackage{geometry}
\usepackage{fancyhdr}
\usepackage{textcomp}
\usepackage{amsmath}
\usepackage{graphicx}
\usepackage{subscript}

\makeatletter

\newcommand{\lyxdot}{.}

\newcommand{\lyxaddress}[1]{
	\par {\raggedright #1
	\vspace{1.4em}
	\noindent\par}
}

\@ifundefined{showcaptionsetup}{}{%
 \PassOptionsToPackage{caption=false}{subfig}}
\usepackage{subfig}
\makeatother

\begin{document}
\title{Time and space resolved first order optical interference between distinguishable
photon paths.}
\author{M. Fernandez-Guasti, C. García Guerrero.}
\maketitle

\lyxaddress{Lab. de Óptica Cuántica, Depto. de Física,\\
Universidad Autónoma Metropolitana - Iztapalapa,\\
C.P. 09340 Ciudad de México, Ap. postal. 55-534, MEXICO\\
e-mail: mfg@xanum.uam.mx, url: https://luz.izt.uam.mx}




\begin{abstract}
Interference between different photons occurs and has been observed
under diverse experimental conditions. A necessary condition in order
to obtain interference fringes is the existence of at least two possible
paths and unknown which-path information. If the photon beams have
different frequencies, stability of the sources and fast enough detectors
are also required.\\
First order interference between two truly independent CW laser sources
is observed. Contrary to what is expected, interference is observed
although the photon beams are distinguishable and the path is unequivocally
known for each photon beam. Segments of the CW wavetrains are selected
with an acousto optic modulator. Temporal and spatial interference
are integrated in a single combined phenomenon via streak camera detection.
The fringes displacement in the time-space interferograms reveal the
trajectories of the labeled photons. These results indicate that in
non-degenerate frequency schemes, the ontology has to be refined and
the which path criterion must be precisely stated. If reference is
made to the frequency labeled photons, the path of each photon is
known, whereas if the query is stated in terms of the detected photons,
the path is unknown.
\end{abstract}

\section{Introduction}

The two manifestations of first order interference between two wave-fields
are i) spatial interference fringes and ii) temporal interference
or beating. The beams in optical spatial interference setups are usually
derived from the same source but traveling along separate paths. In
contrast, temporal interference is commonly achieved with different
sources having different frequencies. Early experiments in the laser
era demonstrated the temporal \cite{javan1962} and spatial \cite{mandel1963}
interference of independent sources. The position-momentum uncertainty
provided an explanation of these results without having to renounce
to the Dirac statement that 'Interference between different photons
never occur' \cite{Pfleegor67}. However, later experiments where
the photon source statistics were carefully controlled, showed ineluctably
that independent photons can interfere \cite{kaltenbaek2006}. Interference
fringes in the visible region of the spectrum with two different frequencies
have also been observed \cite{bratescu1981,Lee2003}. For a constant
frequency difference, a single laser source is commonly used. The
frequency of one beam can then be shifted with an acousto optic modulator
\cite{Davis1988} or by selecting different frequencies from a spatially
chirped femtosecond pulsed source \cite{Garcia2002a}. If separate
lasers are used, the frequency difference varies from shot to shot
and so does the fringe pattern \cite{Louradour93}. The fringes in
most of these experiments with different frequencies have been observed
using a streak camera and more recently, with a modulated CMOS camera
\cite{Patel2014}.

First order interference involves correlations between the fields
whereas second order interference involves correlations between the
fields' intensities. These correlations can be described with semi-classical
(continuum) field theory (CFT) or quantum field theory (QFT). The
archetypal Young's two slit interference experiment exhibits the same
first order interference patterns when produced by short exposure
intense light sources or by long exposure accumulation with feeble
light. However, the ontology in the two theories is quite different.
CFT invokes a well defined amplitude and phase correlation between
the two interfering fields during the detector integration time \cite{Marathay82}.
In QFT, interference takes place only when the path of the photons
is unknown \cite{Eichmann93}. Recall that the two theories do produce
measurable differences in second order interference experiments \cite{Riedmatten03}.
Many theoretical predictions and experimental verifications without
classical analogue, favor the quantum nature of electromagnetic fields. 

The present experimental results are at odds with the standard formulation
of the quantum which path problem: ``A measurement which shows whether
the photon passed through A or through B perturbs the state of the
photon to such an extent that no interference fringes are detected.
Thus, either we know which slit the photon passed through, or we observe
interference fringes. We cannot achieve both goals: the two possibilities
are incompatible \cite[p.22]{esposito2014}''. Many authors consider
that the which path information problem is an example of Bohr’s principle
of complementarity, where the interference/which-way duality is a
manifestation of the wave/particle mutually exclusive concepts \cite{Jacques2008}.
However, Bohr’s idea of complementarity is a much broader principle
dealing with observation and the definition of quantum states \cite{bohr2011,Plotnitsky2013}.
As we shall presently show, interference with which-path certainty
is possible in non-degenerate frequency schemes if the statement is
made in terms of the frequency labeled photons but without reference
to a detected photon. However, the path of a detected photon in the
interference region, cannot be traced back. This latter, more precise
assertion, is consistent with the prevailing quantum viewpoint. Our
observations are consistent with Heisenberg's uncertainty principle
and Busch measurement/disturbance theorem. They are also consistent
with Englert, which way detector inequality. However, they compromise
certain versions of Bohr's complementary principle.

\section{Experimental considerations and setup}

\begin{figure}[h]
\begin{centering}
\includegraphics[scale=0.71]{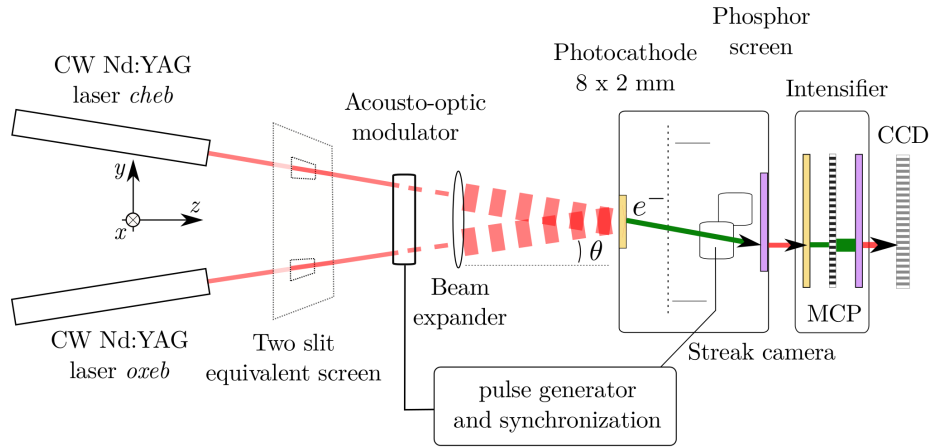}
\par\end{centering}
\caption{\label{fig:arreglo-experimental}Schematic diagram of the experimental
arrangement. The setup is equivalent to a Young's two slit experiment
but each 'slit' is illuminated by an independent laser. The slits
can be conceived to be placed at any plane between the sources and
the photocathode detector before the beams overlap. (Optical beams
drawn in red, electron beams within the streak camera drawn in green.) }
\end{figure}

Two photon beams were generated from two independent Nd:YAG lasers
code named \emph{cheb} and \emph{oxeb} \footnote{Tseltal variant of Mayan language for numbers two $\rightarrow$ \emph{cheb}
and three $\rightarrow$ \emph{oxeb}.}. These continuous wave (CW), monomode lasers (AOTK 532Q) have a
coherence time greater than 300 ns \cite{fernandez-guasti2011a}.
The operation of each laser does not rely in any way on the working
of the other laser, nor are they synchronized. The temperature of
each of them was monitored and controlled independently. Temperature
was measured with a 100 $\Omega$ platinum resistance and controlled
with a Peltier module external to the cavity but attached to its base.
A temperature controller (Stanford Research SRC10) provided the electronic
feedback to maintain a stable temperature within 0.01 °C. The wavelengths
of the two lasers, measured with an spectrometer (Spex1704 with 0.01
nm resolution), were temperature tuned so that their frequencies were
sufficiently close to be resolved by a streak camera (Optronis SC-10).
At sweep speeds of 10 ns/mm with the TSU-12-10 unit, fringes are comfortably
observed in the streak camera for frequency differences below 1GHz.
The two laser beams were steered with mirrors into a TeO\textsubscript{2}
acousto optic modulator (AOM) (10/10 ns, 10 to 90\% rise/fall time
for a 55 $\mu m$ beam-waist). The general setup is shown in figure
\ref{fig:arreglo-experimental}. Preliminary results were reported
at a PIERS conference \cite{Fernandezguasti2020pb}. Beam splitters
were avoided (except for alignment purposes, prior to operation),
so that a two slit wavefront division interferometer is emulated throughout
the trajectories. The expanded collimated beams were overlapped and
detected with the streak camera. The beams collimation was adjusted
with the aid of a shear interferometer. In the streak camera, also
called optical oscilloscope, light impinges on a photocathode placed
on the inner part of a vacuum tube. The electrons emitted by the photocathode
(8 mm x 2 mm) are accelerated and swept in the perpendicular direction
to its long axis, so that a two dimensional image is produced. Each
photoelectron thereafter impacts on a multichannel plate (MCP) and
is cascaded so that the bunch of electrons produces a bright point
 as it reaches a phosphor screen. The MCP amplification voltage is
adjusted so that the intensity of this spot is adequately detected
by a CCD camera. At low intensity levels, the device can be operated
in photon counting mode. At some spots more than one photon is detected,
this information is encoded onto a level of gray, typically not more
than 50 events per 0.7 ns as can be seen from figure \ref{fig:Oxeb-blue-spat1}b.
In the streak camera plots, the abscissa represents time whereas the
ordinate represents a transverse spatial coordinate. The density of
white spots is proportional to the photon density in the two dimensional
time-space coordinates. Streak images cannot be accumulated in this
setup because the frequency and relative phase between the two lasers
vary stochastically in time at scales larger than the coherence time.
Therefore, the fringe pattern was recorded in single exposures with
time duration of the order of the coherence time. Low repetition rates
between 1 and 3 Hz were used to acquire images in real time. The streak
camera detector performs an integration both in space and time,
\[
\left\langle I\left(t,y\right)\right\rangle _{\delta t,\delta x,\delta y}=\frac{1}{\delta t\delta x\delta y}\int_{y}^{y+\delta y}\int_{x}^{x+\delta x}\int_{t}^{t+\delta t}I\left(\tau,\xi,\eta\right)d\tau d\xi d\eta,
\]
where $I\left(\tau,\xi,\eta\right)$ is the light intensity incident
on the photocathode as a function of time and the transverse dimensions.
$I\left(t,y\right)$ is the intensity at the CCD screen as a function
of the 'coarse grain' time and one spatial direction. In the transverse
dimensions, the $x$ direction is limited to $x=0\pm7.5\:\mu\textrm{m}$
by a $\delta x=15\,\mu$m entrance slit; In the $y$ direction, the
position detection range is 15 mm with $\delta y=70\,\mu$m resolution.
The temporal sweep is performed in the $x$ direction with 0.34 \%
resolution of the full sweep time.  This instrumental integration
together with the image amplification and digitization establishes
the resolution of the apparatus. The low noise photocathode has a
10. 37 \% quantum efficiency (QE) at 532 nm and a dark noise of $100\:e^{-}/\mathrm{c}\mathrm{m}^{2}\mathrm{s}$
(Photek ST-LNS20). Temporal resolution at the phosphor screen is 66
$\mu$m (Anima-PX/25, 19.5 x 14.9 mm\textsuperscript{2}, 12 bit A/D
CCD 1392 $\times$ 1024).

\begin{figure}[h]
\begin{centering}
\subfloat[\label{fig:oxcheb}97 ns pulse from laser code named \emph{oxeb.}]{\begin{centering}
\includegraphics[clip,scale=0.07]{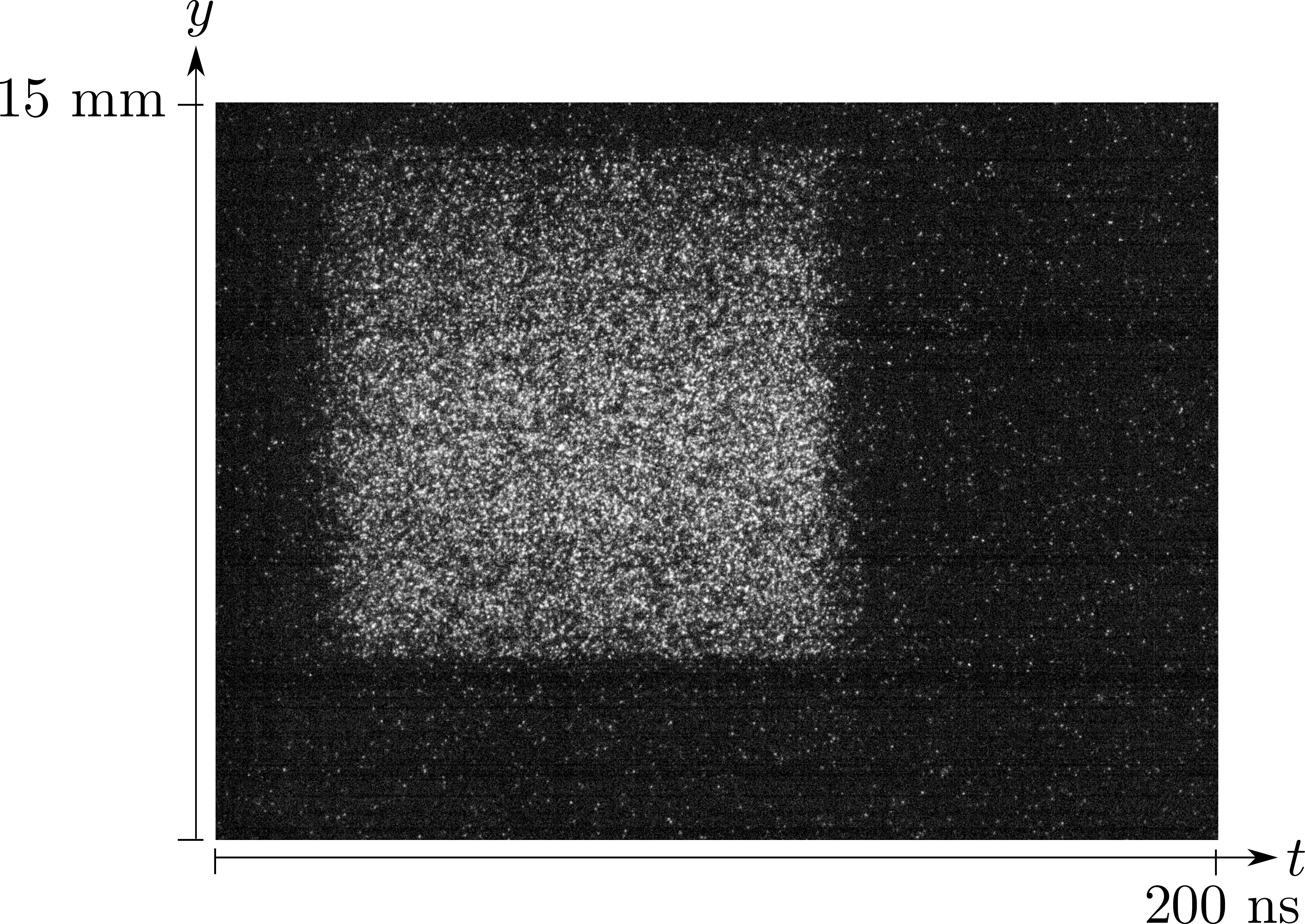}
\par\end{centering}
}\hfill{}\subfloat[\label{fig:Superposition-of-both}Interferogram when both laser pulses
are present.]{\begin{centering}
\includegraphics[clip,scale=0.07]{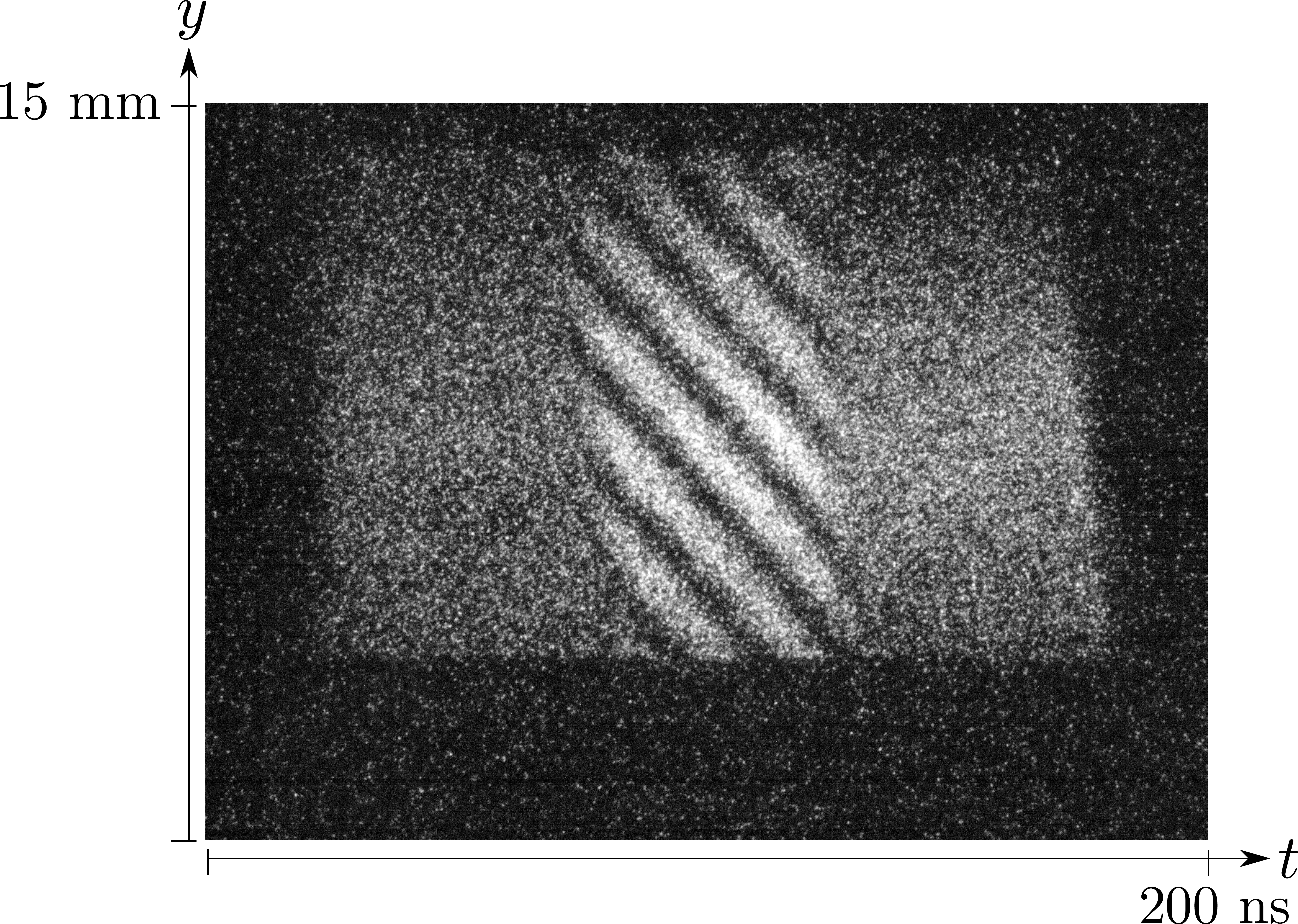}
\par\end{centering}
}\hfill{}\subfloat[\label{fig:cheb}95 ns pulse from laser code named \emph{cheb.}]{\begin{centering}
\includegraphics[clip,scale=0.07]{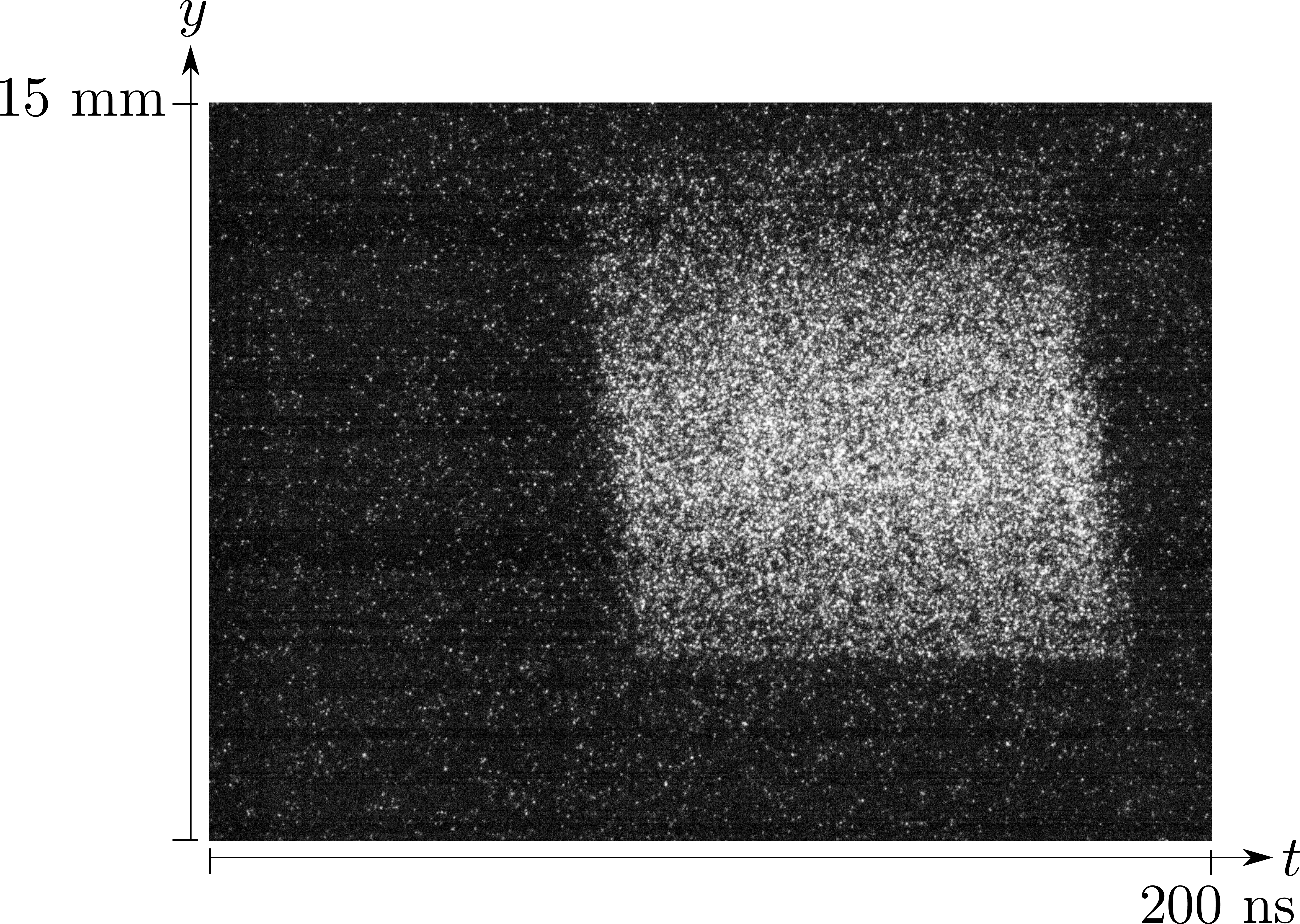}
\par\end{centering}
}
\par\end{centering}
\caption{\label{fig:streak-camera-images-2}Streak camera images. The abscissa
represents time at 10 ns/mm sweep rate with $\delta t=1$ ns temporal
resolution. The ordinate depicts the beams transverse distance in
the $y$ direction.}
\end{figure}

The AOM input angle was aligned with the oxeb laser beam. In our experiments,
delay generator pulses with 120, 300 and 700 ns width were used. The
AOM first order deflection angle is 25 mrad with a diffraction frequency
shift of 210 MHz. The oxeb laser beam was diffracted in first order
(210 MHz) whereas the cheb beam was operated in second order (2$\times$210
MHz). There was thus a 210 MHz frequency difference that posed no
problem because it was compensated by the laser's temperature tuning.
Half wave retardation plates were used at the output of each laser
to adjust the polarization plane in order to improve fringe visibility.
A typical $\approx$100 ns output pulse is shown in figure \ref{fig:oxcheb}.
The cheb beam is a bit mismatched due to a 25 mrad  oblique incidence
in a displaced region of the acoustic wave within the crystal. This
inclination produces a pulse delay and a pulse front tilt as may be
seen in figure \ref{fig:cheb}. This sweep mode neatly exhibits three
regions, two to the left and right in figure \ref{fig:Superposition-of-both},
where only one of the pulses is present and the central area where
both pulses overlap in time and interference is observed.

In a conventional single source Young's experiment, the slits separate
the wavefront into two distinct wavefronts. Here, the wavefronts emanate
from two different sources. The physical setup can be conceived as
each of the sources illuminating a slit. The plane where the slits
are placed is any plane between the sources and the plane where the
beams begin to overlap just before the interfering detection plane.
Furthermore, each slit could be placed at a different plane, reminiscent
of second order interference patterns produced by non-local objects
\cite{vidal2008}. 

\subsection{Photon  labeling}

The two collimated beams are incident on the streak camera photocathode
with a 0.14 mrad angle between them in order to have comfortably resolved
 fringe maxima separated by 1.88 mm.  Cartesian coordinates are
set with $z$ normal to the detector surface and the fields are linearly
polarized in the $x$ direction. The two wave-fields propagate in
the $\left(y,z\right)$ plane, paraxially to the $z$ direction at
a small but opposite angle $\theta$ in the $y$ axis. The wave vector
of the field coming from the cheb laser is
\[
\mathbf{k}_{1}=\mathbf{k}_{y1}+\mathbf{k}_{z1}=-k_{1}\sin\theta\,\hat{\mathbf{e}}_{y}+k_{1}\cos\theta\,\hat{\mathbf{e}}_{z},
\]
and the wavefield coming from the oxeb laser is
\[
\mathbf{k}_{2}=\mathbf{k}_{y2}+\mathbf{k}_{z2}=k_{2}\sin\theta\,\hat{\mathbf{e}}_{y}+k_{2}\cos\theta\,\hat{\mathbf{e}}_{z},
\]
where $\left|\mathbf{k}_{1}\right|=k_{1}=\frac{\omega_{1}}{c}$ and
$\left|\mathbf{k}_{2}\right|=k_{2}=\frac{\omega_{2}}{c}$ are the
respective wave vector magnitudes and $\hat{\mathbf{e}}_{y},\hat{\mathbf{e}}_{z}$
are unit vectors in the $y$ and $z$ directions. The $y$ axis positive
direction has been set in the same direction of $\mathbf{k}_{y2}$,
that is, the photons coming from the oxeb laser have positive momentum,
$\hbar\mathbf{k}_{y2}=\hbar\left|\mathbf{k}_{y2}\right|\hat{\mathbf{e}}_{y}$
 at the detector plane (The $y$ axis positive direction could have
been set in the opposite sense. Either convention applied consistently
yields the same results). In addition, the photons are also labeled
by their frequency. Since each laser source has its specific frequency,
the wavefield from the \emph{oxeb} laser with momentum projection
$\hbar\mathbf{k}_{y2}$ has frequency $\omega_{2}$ and the wavefield
from the \emph{cheb} laser with momentum projection $\hbar\mathbf{k}_{y1}$
has frequency $\omega_{1}$. The wave vector projection in the transverse
$y$ direction and the corresponding frequency are highly correlated.
This so-called labeling of the photons is similar to the temporal
and spatial labeling terminology in HOM second order interferometers
\cite{lee2006}. However, the frequency of each laser is not known
a priori, due to the fluctuations (albeit tiny) in the two lasers.
As we shall presently see, it is only when a set of quantum tests
is performed that the relative frequencies of the two lasers can be
inferred.

\subsection{\label{subsec:QFT-description}QFT description}

In QFT, the standard representation of two quantized complex electric
field operators with linear polarization is $\hat{E}_{1}^{\left(+\right)}\left(\mathbf{r}\right)=i\mathcal{E}_{1}^{\left(1\right)}\exp\left(i\,\mathbf{k}_{1}\cdot\mathbf{r}\right)\hat{a}_{1}$
and $\hat{E}_{2}^{\left(+\right)}\left(\mathbf{r}\right)=i\mathcal{E}_{2}^{\left(1\right)}\exp\left(i\,\mathbf{k}_{2}\cdot\mathbf{r}\right)\hat{a}_{2}$,
where $\mathcal{E}_{1}^{\left(1\right)},\,\mathcal{E}_{2}^{\left(1\right)}$
are the one-photon amplitudes and $\hat{a}_{1},\,\hat{a}_{2}$ are
the annihilation operators for modes 1 and 2 respectively. The fields
from each monomode laser are adequately represented by single mode
coherent states $\left|\alpha_{1}\right\rangle $ and $\left|\alpha_{2}\right\rangle $
\cite{aspect2010}. These quasi-classical states are eigenstates of
the annihilation operators $\hat{a}_{1}\left|\alpha_{1}\right\rangle =\alpha_{1}\left|\alpha_{1}\right\rangle $
and $\hat{a}_{2}\left|\alpha_{2}\right\rangle =\alpha_{2}\left|\alpha_{2}\right\rangle $
with eigenvalues $\alpha_{1},\,\alpha_{2}$. Since the two fields
are independent, their superposition is a two mode factorizable state,
$\left|\psi_{1,2-qc}\left(t\right)\right\rangle =\left|\alpha_{1}\exp\left(-i\omega_{1}t\right)\right\rangle \left|\alpha_{2}\exp\left(-i\omega_{2}t\right)\right\rangle .$
These states allow for the factorization of the first order coherence
function \cite{glauber2007}. The quantum photo detection probability
is $w\left(\mathbf{r},t\right)=s\left\langle \psi_{1,2-qc}\left(t\right)\right|\hat{E}^{\left(-\right)}\left(\mathbf{r}\right)\hat{E}^{\left(+\right)}\left(\mathbf{r}\right)\left|\psi_{1,2-qc}\left(t\right)\right\rangle $,
where $s$ is the sensitivity of the detector and the operator $\hat{E}^{\left(-\right)}\left(\mathbf{r}\right)$
is the Hermitian conjugate of the positive frequency part of the electric
field operator $\hat{E}^{\left(+\right)}\left(\mathbf{r}\right)$.
This expression evaluates to
\begin{multline}
w\left(\mathbf{r},t\right)=s\left(\mathcal{E}_{1}^{\left(1\right)}\mathcal{E}_{2}^{\left(1\right)}\right)^{2}\\
\left(\left|\alpha_{1}\right|^{2}+\left|\alpha_{2}\right|^{2}+\alpha_{1}^{*}\alpha_{2}\exp\left[i\left(\left(\mathbf{k}_{2}-\mathbf{k}_{1}\right)\cdot\mathbf{r}-\left(\omega_{2}-\omega_{1}\right)t+\varphi_{2}-\varphi_{1}\right)\right]+c.c.\right),\label{eq:quant photo detect}
\end{multline}
where $\varphi_{1},\,\varphi_{2}$ are independent stochastic functions
with coherence times $\tau_{1},\,\tau_{2}$ due to the laser cavities'
instabilities. Recall that in this experiment, the coherence time
of each laser is somewhere above 300 ns. 

The spatially dependent interference argument is $\left(\mathbf{k}_{2}-\mathbf{k}_{1}\right)\cdot\mathbf{r}=2\bar{k}\sin\theta\;y+\triangle k\cos\theta\;z$,
where $2\bar{k}=k_{1}+k_{2}$, $\triangle k=k_{2}-k_{1}$. The field
is observed at a detector placed at the $z=z_{0}$ plane, thus the
term $\triangle k\cos\theta\,z_{0}$, only adds a constant phase shift.
The phase as a function of time and the transverse distance $y$ is
\begin{equation}
\phi=2\bar{k}\sin\theta\:y+\triangle k\cos\theta\,z_{0}-\triangle\omega\,t+\triangle\varphi,\label{eq:phase-interf-term}
\end{equation}
where $\triangle\omega=\omega_{2}-\omega_{1}$ and $\triangle\varphi=\varphi_{2}-\varphi_{1}$.
When two different frequencies are present, the equiphase surfaces
evolve in time and space. In contrast, wave-fronts in frequency degenerate
setups entail spatial coordinates alone. The velocity of an equiphase
plane, provided that $\triangle\varphi$ varies slowly in time and
space, is
\begin{equation}
\frac{dy}{dt}=\frac{\triangle\omega}{2\bar{k}\sin\theta}.\label{eq:slope}
\end{equation}
The fringes are thus displaced in time with a slope $\frac{dy}{dt}$,
whose sign is determined by the value of $\triangle\omega$.

\section{Experimental results}

\begin{figure}[h]
\begin{centering}
\includegraphics[scale=0.66]{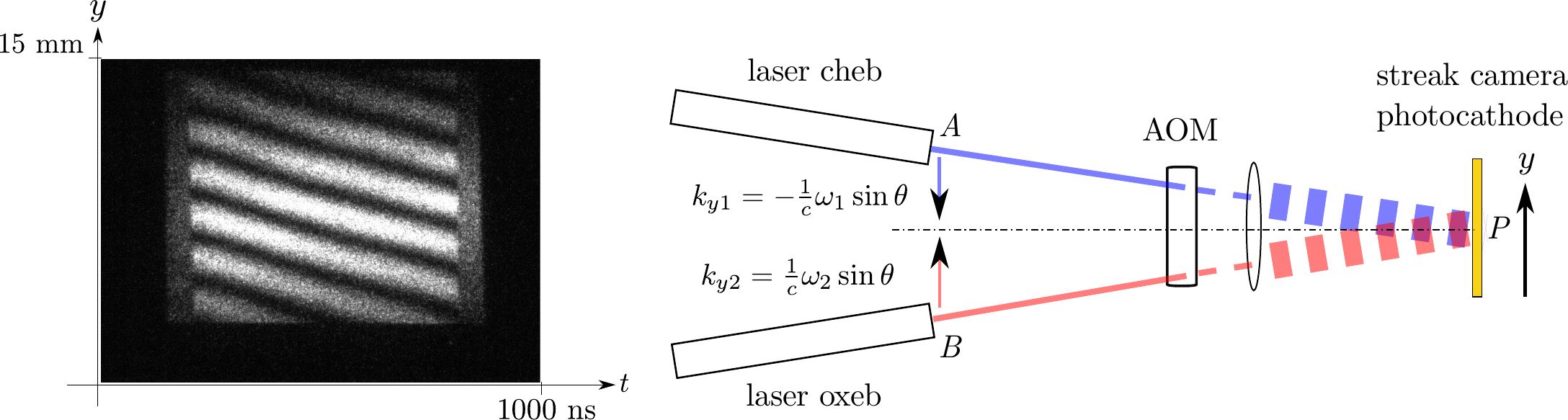}
\par\end{centering}
\caption{\label{fig:Negative-slope-interference}The steak camera image (left)
shows negative slope interference fringes. Therefore, the oxeb laser
emitted lower frequency photons that followed the path $\overline{BP}$.
The higher frequency photons from the cheb laser followed the path
$\overline{AP}$.}
\end{figure}

\begin{figure}[h]
\begin{centering}
\includegraphics[clip,scale=0.66]{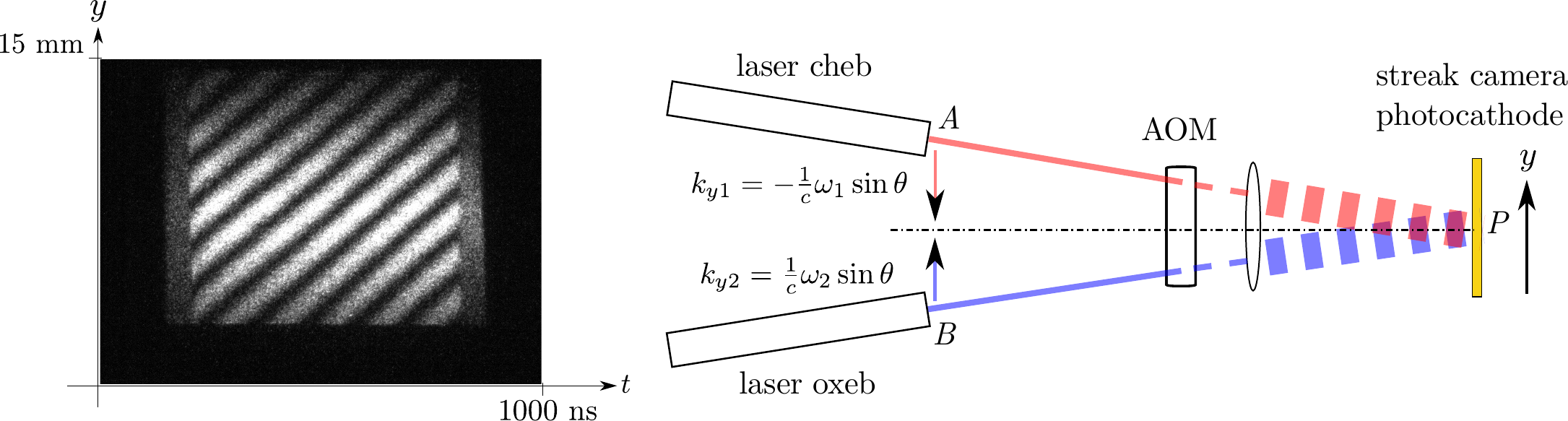}
\par\end{centering}
\caption{\label{fig:Positive-slope-interference}Positive slope interference
fringes, the oxeb laser emitted higher frequency photons than the
cheb laser. Photons from the oxeb laser followed the lower path $\overline{BP}$
whereas photons from the cheb laser followed the upper path $\overline{AP}$.}
\end{figure}

Each point in the streak camera image represents a quantum test of
whether a photon arrived at position $y$ of the streak camera photocathode
at a given time $t$. A streak camera image consists of two sets
of quantum tests, one in the spatial domain and another in the temporal
domain. In the $y$ ordinate direction, electrons in the photocathode
long axis  act as a set of spatially distributed detectors. For
each $y$ position, there is another set of  different consecutive
quantum tests  that probe the dynamical evolution of the quantum
system \cite[p.33, p.237]{peres2002}. This set is depicted in the
abscissas time axis. A photoelectron is emitted if a photon is present
at $\left(y,t\right)$ where the two photon beams overlap with 10.37\%
QE. These events are amplified by the MCP and discretized in the 1024
$\times$ 1392 $=1.425\times10^{6}$ detectors at the CCD. Thus, each
streak camera interferogram involves $10^{6}$ quantum tests (order
of magnitude). The interferograms in figures \ref{fig:Negative-slope-interference}
and \ref{fig:Positive-slope-interference} were registered at 50 ns/mm
sweep speed with 3.4 ns temporal resolution. The transverse spatial
range is 15 mm with $70\,\mu$ resolution. The 655 ns segments obtained
with the AOM modulator, where the two CW lasers temporally overlap,
exhibit high contrast interference fringes with visibility above 70\%.

The slope of the fringes in the interferogram shown in figure \ref{fig:Negative-slope-interference}
is negative. From Eq. \eqref{eq:slope}, if the slope of the equiphase
lines is negative, the frequency difference $\triangle\omega$ is
negative and thus $\omega_{2}<\omega_{1}$. Therefore, the oxeb laser
emitted photons with lower energy $\hbar\omega_{2}$, drawn in red
in figure \ref{fig:Negative-slope-interference}. Each of these photons,
with positive linear momentum projection $\hbar\mathbf{k}_{y2}$ in
the $y$ direction, ineluctably followed the path $\overline{BP}$,
where $B$ is the position of the beam at the laser oxeb output and
$P$ is a point in the streak camera photocathode screen. The converse
is true for the photons that comprise the cheb laser beam. These higher
energy photons drawn in blue in figure \ref{fig:Negative-slope-interference},
have negative linear momentum projection $-\hbar\left|\mathbf{k}_{y1}\right|\hat{\mathbf{e}}_{y}$
 at the detector plane. They followed the path $\overline{AP}$,
where $A$ is the position of the beam at the cheb laser output. From
the time-space interferogram, it is of course possible to evaluate
the frequency difference $\triangle\omega=-19.4$ MHz, although the
specific value is irrelevant for the argumentation. 

The interferogram shown in figure \ref{fig:Positive-slope-interference},
was acquired 1014 ms after the interferogram shown in figure \ref{fig:Negative-slope-interference}.
The lasers' frequency drifted so that the slope changed sign from
one scan to the next. For a positive slope, the frequency difference
$\triangle\omega$ is positive and thus $\omega_{2}>\omega_{1}$.
Therefore, the oxeb laser emitted photons with higher energy $\hbar\omega_{2}$,
drawn in blue in figure \ref{fig:Positive-slope-interference}. Each
of these higher energy photons again necessarily followed the path
$\overline{BP}$. The photons emitted by the cheb laser beam now have
lower energy and followed the path $\overline{AP}$. In this case,
$\triangle\omega=54.9$ MHz. Summing up the two previous results:
\emph{The fringes are displaced, as a function of time, in the same
direction of the transverse momentum projection of the photons with
higher energy}.  It should be stressed that the detected photons
are neither blue nor red but photons with information from both sources
given by the quantum photo detection probability stated in Eq. \eqref{eq:quant photo detect}.

In the interferogram of figure \ref{fig:Positive-slope-interference},
if some red photons had positive momentum (came from the lower slit)
and some blue ones had negative momentum (came from the upper slit),
they would produce fringes with a negative slope. However, the interferogram
in figure \ref{fig:Positive-slope-interference} does not exhibit
even the faintest fringes with negative slope, thus this possibility
is ruled out.  Therefore, we must conclude that in either case ($\omega_{2}>\omega_{1}$
or $\omega_{2}<\omega_{1}$), the trajectory of the photons is well
defined, yet a high contrast interference pattern is observed. 
\begin{figure}[h]
\begin{centering}
\includegraphics[clip,scale=0.66]{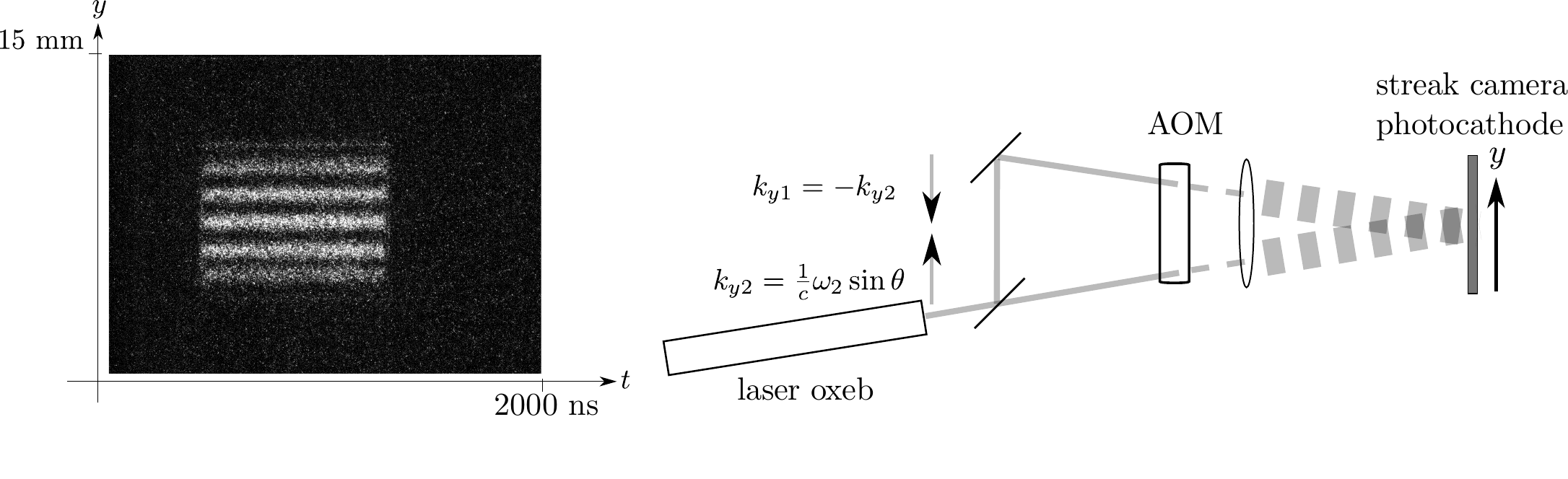}
\par\end{centering}
\caption{\label{fig:Zero-slope-interference}Zero slope interference fringes.
Photons are no longer frequency distinguished.}
\end{figure}

Consider for completeness, the case where the frequencies are equal.
This condition is easier to achieve experimentally using the same
laser source but it could actually be accomplished with two laser
sources with the appropriate stability and bandwidth. In this case,
interference fringes have zero slope and the pattern is constant in
time as shown in figure \ref{fig:Zero-slope-interference}. There
is no frequency labeling of the photons and it is not possible to
deduce which path they followed. There is still a momentum labeling
but, due to the position-momentum uncertainty, the sources are unresolved
at the detector \cite{Mandel83}. 

\section{\label{sec:Evaluation-of-quantum}Evaluation of quantum uncertainties}

In order to establish the photons path, it is sufficient to measure
whether the fringes displacement is positive or negative. Nonetheless,
it is reassuring to confirm that the actual numerical values of the
measurements do not violate an uncertainty relationship, nor are they
buried below the quantum noise.
\begin{figure}[h]
\subfloat[Rectangle width is 0.7 ns (width  is over-sized in image for clarity),
height encompasses the entire $y$ range.]{\includegraphics[scale=0.42]{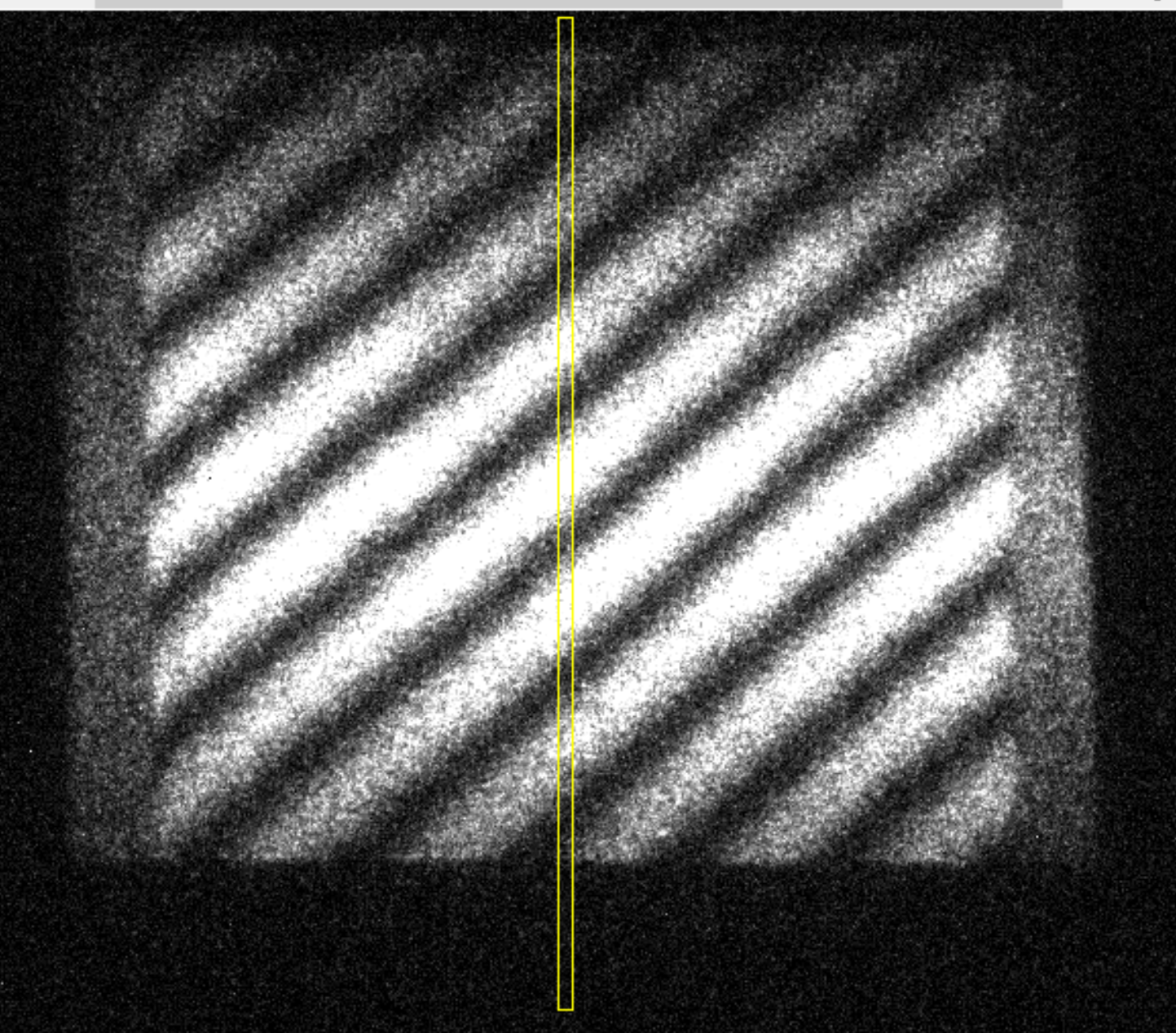}

}\hfill{}\subfloat[Plot of spatial profile averaged over $\triangle t=0.7$ ns (Optoanalyze
v3.71)]{\includegraphics[viewport=0bp 8.98693bp 398.705bp 275bp,clip,scale=0.61]{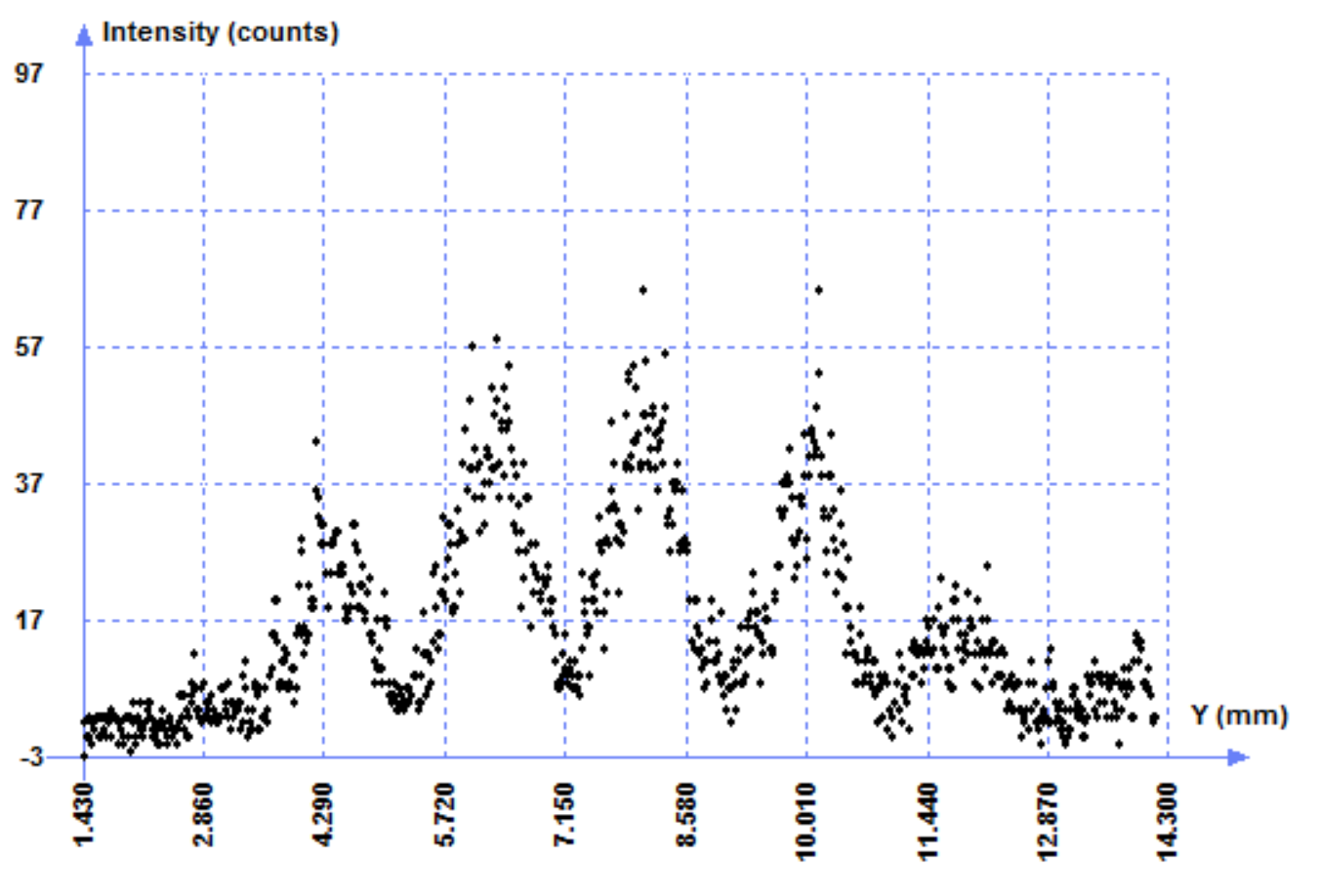}

}

\caption{\label{fig:Oxeb-blue-spat1}Region of interferogram shown in figure
\ref{fig:Positive-slope-interference}, limited to small range in
$t$ but the whole range in $y$.}
\end{figure}

The number of photons $\Phi^{\textrm{ph}}$ per unit time is given
by the power $\Phi$ over the energy per photon $\hbar\omega_{\ell}$.
The power of each laser is 50 mW but losses due to off optimal temperature
operation and beam steering reduces the power roughly by a factor
of 5. For a 10 mW average power with angular frequency $\omega_{\ell}=3.54\times10^{15}\textrm{ Hz}$,
the average number of photons per nanosecond is $\Phi^{\textrm{ph}}=\frac{\Phi}{\hbar\omega_{\ell}}=2.68\times10^{7}\textrm{ photons}\cdot\mathrm{ns}^{-1}$.
The $\delta x=15\,\mu$ horizontal entrance slit reduces the power
by a factor of $10^{-3}$ and the QE of the photodetector by $1.037\times10^{-1}$.
The average number of photons detected per nanosecond is then $\Phi^{\textrm{ph}}=2.68\times10^{3}\textrm{ photons}\cdot\mathrm{ns}^{-1}$.
The standard deviation in the number of photons is thus $\sqrt{\left\langle N_{\ell}\right\rangle }=\sqrt{\Phi^{\textrm{ph}}}=\sqrt{2.68\times10^{3}}=51.7$.
The phase uncertainty in the standard quantum limit (SQL) \cite{clerk2010}
in one nanosecond is then approximately
\begin{equation}
\triangle\phi_{\mathrm{SQL}}=\frac{1}{2\sqrt{\left\langle N_{\ell}\right\rangle }}=9.66\times10^{-3}\approx10^{-2}\textrm{ radians}.\label{eq:sql}
\end{equation}
This value of the SQL establishes the minimum achievable uncertainty
in the phase of each photon beam at the detector. On the other hand,
let us assess the measurement error in the interference pattern produced
by the two sources. From figure \ref{fig:Oxeb-blue-spat1}, the distance
between maxima is  $\triangle y_{max}=1.88\pm0.023\textrm{ mm}$.
The main error coming from the instrumental resolution in the $y$
direction. The spatial resolution between maxima located 1.88 mm apart,
due to the sources' phase uncertainty $\triangle\phi_{\mathrm{SQL}}$
 per nanosecond is $3\,\mu\textrm{m}$. The spatial frequency measurement
uncertainty due to the interference of the two photon beams is thus
about 8 times larger than the SQL of each laser source.

\begin{figure}[h]
\begin{centering}
\includegraphics[scale=0.17]{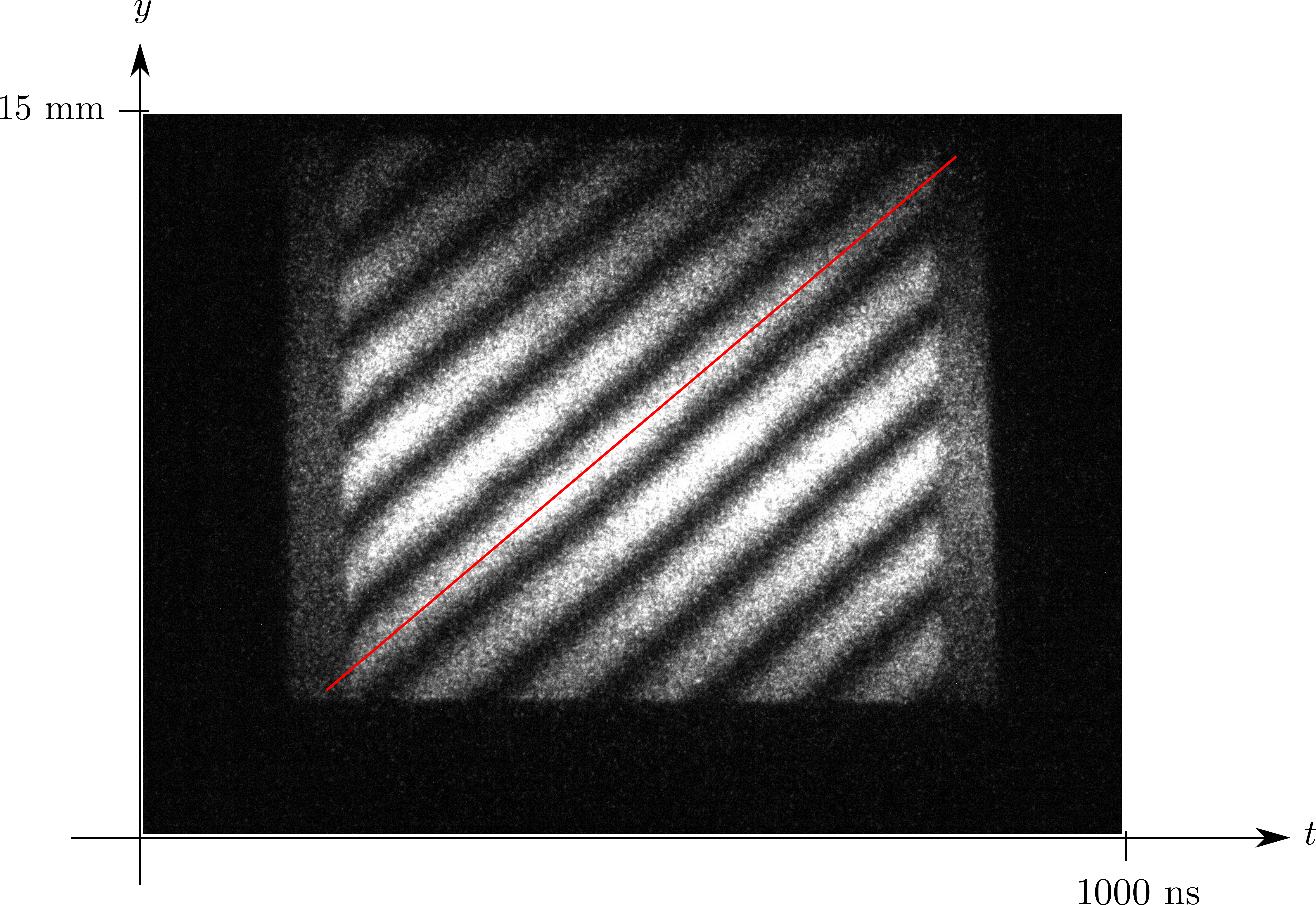}
\par\end{centering}
\caption{\label{fig:slope}Straight line superimposed on the interference fringes
shown in figure \ref{fig:Positive-slope-interference}.}
\end{figure}

The fringes observed in the various interferograms presented in the
paper follow straight lines, thus their slope is constant as predicted
by Eq. \eqref{eq:slope}. Figure \ref{fig:slope}, reproduces the
positive slope interference fringes, (where the oxeb laser emits higher
frequency photons than the cheb laser) together with a straight line
fit. Therefore the lasers relative phase fluctuation $\triangle\varphi$
must be constant (or at most linear in time) during the $\approx$
603 ns where there is temporal overlap of the two beams. Frequency
fluctuations are thus smaller than the long term average $\triangle\nu=$
3 MHz laser linewidth in the $\mu$s timescale. For short detection
times, the laser cavity fluctuations are 'frozen' and the lasers linewidth
approaches the Schawlow-Townes quantum limit, $\triangle\nu_{laser}=\frac{4\pi\hbar\omega}{\tau_{cav}^{2}P}$
\cite{townes1958}. For these Nd:YAG monomode laser systems, the quantum
limit linewidth is of the order of a few kHz.

The existence of an energy-time uncertainty relation has been subject
to much debate \cite{Busch1990,Miyadera2016}. Due to the lack of
a self-adjoint time operator, there is no quantum uncertainty relationship
of time with any other dynamical variable \cite{bohm1961}, in particular
energy or linear momentum. Nonetheless, time and frequency are, of
course, Fourier transform conjugate variables subject to the inequality,
$\delta t\,\delta\omega\geq\frac{1}{2}$ for Gaussian pulses based
on the mean square deviation \cite[p.623]{diels+rudolph2006}. For
example, the beat frequency ($\triangle\omega=54.9$ MHz) in the interferogram
shown in figure \ref{fig:slope}, is obtained from the measurement
during the beam's temporal overlap of $\delta t=603$ ns. The frequency
resolution is thus at most, $\delta\omega\geq1/\left(2\delta t\right)=0.83$
MHz. In quantum parlance, photons in different modes are distinguishable
if the detection time is longer than the inverse of the modes frequency
separation. Photons in modes separated by $\triangle\omega=54.9$
MHz are distinguishable if they are detected in times longer than
$1/\triangle\omega\approx18$ ns.  This time is considerably larger
than the photolelectric response time \cite{liuj2015} and the uncertainty
in the time axis for a single temporal event. Nonetheless, the detection
is performed in successive time measurements over a time span larger
than $603$ ns.

In the limit of macroscopic fields and small quantum fluctuations,
photon number $N_{\ell}$ and phase $\phi_{\ell}$ fluctuations ($\ell=1,2$),
look like complementary variables in the usual sense of quantum mechanics
$\triangle N_{\ell}\triangle\phi_{\ell}\geq\frac{1}{2}$ \cite[p. 368]{aspect2010}.
For minimum uncertainty states and in particular for coherent states,
the equality is fulfilled (fluctuations are proportional to the square
root of the average number of particles in a Poisson distribution)
\begin{equation}
\triangle\phi_{\ell}=\frac{1}{2\sqrt{\left\langle N_{\ell}\right\rangle }}.\label{eq:mus}
\end{equation}
The time-space interferograms shown here nicely depict the trend of
this behavior. For a few scattered dots, $\left\langle N_{\ell}\right\rangle $
is small and constant phase lines are difficult to establish. However,
as the number of dots increase, the equiphase lines become better
delineated and their uncertainty is thus reduced. The number of detected
events $N_{\ell}$ can be varied, either by attenuation of the sources
or by evaluation of a limited portion of the interferogram. In the
latter case, if the slope is evaluated from a partial region of the
image, the number of dots is smaller and the uncertainty in the phase
(slope) becomes larger.

\section{Experimental rationale}

Many which-path experiments have been tried out: ``A succession of
suggestions for more and more ingenious experiments has failed to
provide any method for simultaneous fringe and path observations''
\cite[R. Loudon, What is a photon?]{roychoudhuri2003}. So called
welcher weg experiments were even proposed by eminent physicists,
Einstein and Feynman amongst them \cite[Sec.1.1.3]{ficek2005}. The
less disruptive probes implemented so far involve weak measurements
that provide fuzzy quantum information \cite{aharonov1988,Kocsis2011}.
Our setup, designed to study the dynamics of decoherence, was not
intended to undertake a which-path problem; The before mentioned facts
being enough to deter almost anyone from doing so. Nonetheless, we
should also mention that interference experiments with photons of
different energy were already indicative of a well-known frequency
going to a specific slit \cite{Garcia2002,grave2010}.

Why then does this experiment succeed in the measurement of path knowledge
without destroying the interference pattern? From our understanding,
there are three reasons:
\begin{enumerate}
\item \emph{The path information is obtained from measurements at the interference
detection plane.}
\begin{enumerate}
\item The trajectory is in no way perturbed since the path detection is
not performed in mid trajectory but at the end plane where the fringes
are observed. The Englert inequality establishes that for a given
fringe visibility there is an upper bound on the amount of information
that can be stored in a which-way detector (WWD) \cite{englert1996}.
Englert inequality is derived assuming that the WWD's are placed somewhere
in the way between the two alternative trajectories before the photon
beams overlap. Here, the photocathode plays the role of the WWD's;
However, it is placed at the interference plane where the beams overlap
but not before.
\item Recall that no information can be obtained without disturbing a quantum
system \cite[P. Busch, Quantum Limitations of Measurement]{Myrvold2009}.
In the present experiment, photons are destroyed when detected at
the streak camera photo-cathode where information is extracted, thus
Busch theorem is not violated. Our measurement is not a weak measurement.
On the contrary, each of the $10^{6}$ quantum tests of a given frame,
destroy the photons involved in each test. The system is destroyed,
that is completely disturbed, by the measurement. 
\end{enumerate}
\item \emph{The fringes slope in the time-space coordinates is the decisive
parameter in order to establish the photons path.} 
\begin{enumerate}
\item \label{enu:sufficiently-large}A sufficiently large number of photons
need to accumulate to produce a fringe pattern. Whether this pattern
is obtained by intense or attenuated beam exposures does not alter
the statistics of the laser light and are thus entirely equivalent
\cite{kaltenbaek2006}. It does not make sense to ask whether a single
photon produces a fringe pattern. Nonetheless, the collection of measurements
 gives information about each trial even to the point of stating
that ``Each photon then interferes only with itself'' \cite[p.9]{dirac1978}.
In an analogous fashion, the trajectory of the photons is revealed
here from the measurement of a large number of events. Nonetheless,
information about the trajectory of each photon is obtained.
\item \label{enu:Successive-time-measurements}Successive time measurements
of the fringe pattern are recorded. This scheme follows the rationale
of quantum measurements distributed in time where the path-integral
formulation is particularly well suited to describe the problem \cite{caves1986}.
Feynman's rules for combining probability amplitudes depend on whether
intermediate states are measured \cite{feynman1948}. In the present
experiment no intermediate state is measured. Nonetheless, information
about intermediate states is obtained from measurements at a succession
of final states. 
\end{enumerate}
\item \emph{Photons need to be frequency labeled.} \\
In general, photons need to be doubly labeled with tags that are not
conjugate variables. In this experiment, labels are 'photon linear
momentum projection in the $y$ axis' and 'photon energy' or quantities
derived thereof. Thus determination of one of them does not obstruct
the determination of the other. 
\end{enumerate}
Regarding point \ref{enu:sufficiently-large}, it could be argued
that only the average behavior of the system is being probed. However,
this is not the case. In the prevailing Copenhagen view of quantum
mechanics, or its modern quantum Bayesian version, the theory is intrinsically
probabilistic. A prediction can only be related to observation in
an statistical way given by Born's rule. The larger the number of
measured events, the sharper the measured property (within the uncertainty
principle if complementary variables are involved). From the measurement
of a large number of independent events, it is possible to infer certain
properties of each event. The fundamental reason being that events
independence imply that each event is not altered in any way by
the other events. 

The uncertainty principle has been stated as ``Any determination
of the alternative taken by a process capable of following more than
one alternative destroys the interference between alternatives''
\cite[1-2, p.9]{feynman2010}. This assertion by Feynman and coauthors
is certainly compromised by the present results.  However, they do
not contradict the uncertainty principle. Heisenberg's uncertainty
principle is, strictly speaking, related to the uncertainty between
conjugate variables, that is, operators that do not commute \cite{peres2002}.
In section \ref{sec:Evaluation-of-quantum}, we have shown that the
present experimental results are in full accordance with quantum uncertainties.

\section{Ontology and Discussion}

\subsection{Which way query}

In order to clarify the delicate conceptual difference of the which
path query, let us pose two questions that are seemingly the same
but have different answers:
\begin{itemize}
\item Do the experimental results reveal which path each photon followed?
\end{itemize}
The answer is YES. Let the outcome of the $10^{6}$ quantum tests
be positive slope fringes. Then, each red photon came through A and
each blue photon came through B. The path that each photon followed
is known, yet, an interference pattern is observed. The interference
pattern is built up by the accumulation single photon events. The
certainty of the assertion depends on the visibility of the interference
fringes, and these in turn, depend on the number of quantum events
(and of course, the appropriate experimental arrangement with truly
independent but stable enough sources).
\begin{itemize}
\item Do the experimental results reveal which path did a detected photon
(a white speck on the screen) followed?
\end{itemize}
The answer is NO. When we refer to 'this' photon that impinged on
the screen, it is not known whether it is a red or a blue photon or
even a redblue photon. In order to specify which way it followed,
the color must be known but we only detect a white speck regardless
of the photon frequency.  Thus interference is observed but the detected
photon path is unknown.

The subtle but fundamental difference between these two queries
is that the former question does not involve the category of the detected
entity. In contrast, the detected entity is at the core of the latter
question. 

\subsection{Detected photons}

A closely related but different question is the nature of the detected
photons. Before embarking onto it, we should be aware of the tacit
assumption that photons are considered to exist as an indivisible
lump of electromagnetic energy or at least provide the best description
we have so far of the EM fields. To some extent, this is a matter,
as Penrose puts it, of quantum faith \cite{penrose2016}. A faith
not exempt of support and vast evidence considering the overwhelming
success of quantum field theory. For this reason, the alternatives
mentioned here below do not admit the possibility of a detector (say
an atom) absorbing part of one photon and part of another photon,
for this would destroy the photon concept altogether. 

Two alternatives are envisaged regarding the nature of each detected
photon:

\emph{i) Detected~photons~are~either~blue~or~red.} One possibility
is to consider that a detected photon is either blue or red but its
frequency cannot be known if interference occurs. An asset  of this
approach is that the entities 'red photon' or 'blue photon' retain
their identity. Thus, the photon concept remains a good concept, in
the sense of good quantum numbers. However, this view has the major
problem that there is then no superposition of the disturbances,
but it is superposition that produces the interference phenomenon.
A thought experiment has been proposed before, involving frequency
sensitive photo detectors with different predictions for the expected
outcome \cite[App. A]{Garcia2002}. It has also been stressed that
superposition actually takes place only in the presence of charges
that respond to the superimposed fields \cite{Roychoudhuri2017}. 

\emph{ii) Detected~photons~bear~information~of~both~frequencies
(mainstream view).} The other possibility is to consider that a detected
photon within the interference region has information on both laser
fields as expounded by Paul \cite[p.221]{paul1986}. In the present
experiment, it must bear information of both frequencies according
to the superposition described in subsection \ref{subsec:QFT-description}.
The difficulty with this view is that a photon cannot give part of
it to another photon because it would then loose its entity. Somehow,
it has to give information to the other photon while retaining its
photon identity. Photons cannot be conceived like classical particles.
What is more, photons cannot even be conceived like other quantum
material particles because, in general, there is no mathematical object
that represents a photon wavefunction \cite{scully2001}. The photon
notion arises naturally in number states as the elementary energy
unit $\hbar\omega$. Number states are eigenstates of the Hamiltonian
but their phase is random. In order to observe first order interference,
a well defined phase, up to uncertainty limitations, is required.
Single mode coherent states exhibit a well defined phase but are not
eigenstates of the Hamiltonian. Their energy is not well defined due
to the uncertainty in photon number but, being single mode states,
the energy per photon is fixed. As mentioned by Paul \cite[p.221]{paul1986},
in the detection process, ``an energy packet $h\nu$ is taken from
the superposition field to which both lasers contribute equally, and
hence it is only natural that this photon bears information on both
laser fields''.

\subsection{Final remark}

The which way assertion should be refined in order to have an unambiguous
meaning: The path that each photon followed can be known without destroying
the interference pattern. In this formulation of the statement, the
slit that each photon passed through is known, but it is not known
to which detected photon it corresponds. Another, equally correct
formulation is that, within the interference region, the path of a
detected photon cannot be traced back. That is, if interference occurs,
it is not possible to assert the path followed by a photon detected
on the screen.

\medskip{}
\textbf{Acknowledgements} 

Part of the equipment used in these experiments was funded by CONACYT,
projects CB2005-51345-F-24696 and CB2010-151137-F. 

\medskip{}
On behalf of all authors, the corresponding author states that there
is no conflict of interest.

\end{document}